\documentclass[journal]{IEEEtran}
\usepackage{graphicx,color,braket}
\usepackage{booktabs}
\usepackage{cite}
\usepackage{lipsum}
\ifCLASSINFOpdf
\else
\fi
\usepackage[colorlinks = true]{hyperref}

\begin{document}
	
\title{Analysis of Quantum Key Distribution based  Satellite Communication}
\author{Vishal Sharma{$^{*}$}, Member IEEE Photonics Society and Subhashish Banerjee{$^{*}$} 
\begin{center}{$^{*}$}IIT Jodhpur, Rajasthan, India \end{center}
\begin{center} E-mail:\{pg201383506@iitj.ac.in, subhashish@iitj.ac.in\}
\end{center}
        \IEEEmembership{}
        \IEEEmembership{}
\IEEEcompsocthanksitem }

\markboth{}%
{Shell \MakeLowercase{\textit{et al.}}: Bare Demo of IEEEtran.cls for Journals}
\maketitle

\begin{abstract}
Quantum key distribution is an effective encryption technique which can be used to perform secure quantum communication between satellite and ground stations. Quantum cryptography enhances security in various networks such as optical fibers and wireless networks. In addition to this, these networks become vulnerable in presence of high attenuation due to atmospheric effects and noise. Hence, errors occur due to decoherence. The noisy quantum channel is modeled and implemented by the redundancy-free quantum error correction scheme which provides better security and throughput efficiency as shown in simulation results.
\end{abstract}
\maketitle
 
\section{Introduction}
\IEEEPARstart{I}{n} quantum-based satellite  networks, one of the  central demands for free-space quantum communication is the ability of  successful transmission of qubits under noisy environments \cite{sharma2015controlled, sharma2016comparative, sharma2016effect, sharma2017decoherence, sharma2014analysis, johnson2015analysis,gyongyosi2013pilot, sharma2017analysis}. There are a number of specific techniques which have been used in space applications, for example,  frequency hopping spread spectrum, game theory and routing methods for different spread spectrum approaches \cite{dixon1994spread, han2012game, wright2000wireless}.\newline

Quantum key distribution (QKD) protocols are deployed for satellite communications, but their performance is affected due to environmental noise, adversary attacks, atmospheric turbulence, and telescope dimensions \cite{sharma2017decoherence, sharma2017analysis}. To improve the performance of QKD-based satellite communication under such situations, efficient quantum error correction methods are implemented \cite{wootters1982single, bennett1984quantum}.\newline

A quantum cryptography protocol was proposed in \cite{lutkenhaus1999estimates}, which takes into account on individual information carriers and provides a successful secure key generation. Quantum low-density parity check (LDPC) codes were introduced for addressing the balance between quantum LDPC code performance and entanglement consumption \cite{xie2012channel}. In addition to this, various classical error correction schemes were proposed in \cite{johnson2015analysis}, namely Winnow protocol, LDPC protocol, and Cascade protocol. In these protocols numerous performance comparison parameters such as throughput efficiency, computational, communication complexity, and run time was evaluated \cite{johnson2015analysis}. Further, to enhance the security and quantum throughput efficiency, a scheme of pilot quantum error correction was proposed in \cite{gyongyosi2013pilot}, which was developed specially for quantum based satellite communication where errors introduced due to polarization mismatch.\newline

In this paper, we discuss  error correction schemes and their probabilities, their performance comparison, specially for the quantum-based  satellite communication, including laser repetition rate as a performance index with overall quantum channel estimation. In quantum networks while operating under free space we need to overcome the problems related to laser-beam widening, photon losses due to atmospheric interactions, dark-count rates, background noise, various day-night lightning conditions, and imperfections in photo-detectors. Hence, to improve the performance and security of quantum key distribution based satellite networks under such conditions, proper modeling of quantum channel is required. We then apply an effective quantum error correction technique to improve the quantum throughput efficiency. At the end, we observe from our results that pilot based-quantum error correction scheme performs much better than other already existing error correction methods.\newline

The organization of the paper is as follows. Quantum key distribution based satellite communication and associate problems in performing under non-ideal (noise/attacks) cases are described in section II. Modeling of quantum channel used to perform QKD-based satellite communication networks is elaborated in section III. In addition to this, pilot quantum error correction scheme varying from single qubit to multi-qubit methods are introduced in section IV. The results and conclusions are discussed in sections V,  and VI, respectively.
	    
\section{Quantum Key Distribution (QKD) for satellite networks}

 In classical communication, bit 0 and 1 is the basic unit to represent information. In quantum communication photons and electrons are the fundamental unit to express information carriers which is known as qubits. In mathematical notation these qubits are written in the form of Dirac notation, $|0\rangle$, and $|1\rangle$, which corresponds to classical bits 0 and 1, respectively. There are infinitely many quantum state representations in between $|0\rangle$ and $|1\rangle$, also these quantum states are in the superposition of both $|0\rangle$ and $|1\rangle$ at a time. In classical world, there is no such analogy. In present context, the laser emitted photon's polarization is  represented by these quantum states, $|0\rangle$ and $|1\rangle$, which can be understood as 0 and 1, respectively for classical information. Qubit representation follows 
 
\begin{equation}
|\psi\rangle = \alpha|0\rangle + \beta|1\rangle.
\end{equation}

In the above quantum state representation $\alpha$, and $\beta$ are the complex coefficients and the associate probabilities of outcomes $|0\rangle $ and $|1\rangle $ are $|\alpha|^{2}$,  and $|\beta|^{2}$, respectively. Also, $|\alpha|^{2}$ + $|\beta|^{2}$ = 1.\newline

Quantum key distribution (QKD) shares a secure encryption key between legitimate users (Alice and Bob) \cite{rarity2002ground}. Security of this encrypted key is based on the laws of quantum mechanics. Any eavesdropping attempt is detected by the measurement property of quantum mechanics \cite{wootters1982single}.\newline

A description of quantum key distribution \cite{le2006enhancement} with two channels, quantum and classical are shown in Fig. \ref{fig1}. Initially,  random bits (known as raw key) are generated by Alice. Next, Alice chooses random bases corresponding to each bit, $|H\rangle$, $|V\rangle$, $|\frac{\pi}{4}\rangle$, and  $|\frac{3\pi}{4}\rangle$,  known as conjugate bases. The encoding can be performed by encoding $0$ for $|V\rangle$, and $1$ for $|H\rangle$. In quantum-based satellite networks, after qubit generation, Alice transmits these qubits to Bob via free space as a quantum channel. At receiver side, Bob performs quantum measurements by choosing random conjugate bases.  Here the success probability is $\frac{1}{2}$, because the measurement bases are randomly chosen. After measurement, Bob announces his measurement bases to Alice via a classical channel. Finally a sifted key is generated which corresponds to the same basis used to encrypt and decrypt the random data. This sifted key is the same sequence of bits for both the sender and receiver \cite{hwang2003quantum}.\newline

Quantum-based satellite communications in free-space is a challenging problem. It faces many problems such as turbulence generated losses, geometrical losses due to telescope dimensions, losses due to detector dark count rate, inefficient quantum devices responsible for information leakage, and  disturbances due to eavesdropper attempts \cite{ursin2007entanglement, hughes2002practical, hwang2003quantum}.\newline

Entangled-based quantum communication is an effective and faithful approach which provides better security as compared to faint pulse quantum key distribution technology \cite{ribordy2000long}. Once a proper entanglement distribution is achieved among various desired nodes, a global quantum network is established \cite{sasaki2011field}. In addition to this, as shown in Fig. \ref{fig1}, we need to apply an effective error correction technique to correct the errors obtained from the quantum channel, and finally privacy amplification is applied to get the final secure keys. In general, privacy amplification methods are performed to eliminate eavesdropper generated disturbances \cite{ursin2007entanglement}. \newline
	    \section{Basic Operations For Quantum-Based Satellite Communications }
	    
	    Here, we will concentrate on quantum error correction methods to eliminate the effects considered earlier. It is essential to improve the quantum throughput efficiency for better performance in real field applications. It can be achieved by applying an effective quantum error correction technique.\newline
We can model the non-ideal quantum channel by  $U_{\theta}$. The error introduced in the  channel due to noisy environment can be written as an angle $\theta$ $\in$ [0, 2$\pi$), which could, for example, mode the angular motion of the satellite. 

\begin{equation}
|\theta\rangle = cos\frac{\theta}{2}|0\rangle + i sin\frac{\theta}{2}|1\rangle, \label{eqn1}
\end{equation}

\begin{equation}
U_{\theta} = e^{i \frac{\theta}{2} \left[\begin{array}{cc}
1 & 0\\
0 & -1
\end{array}\right] } 
 \stackrel{\Delta}{=}  cos\frac{\theta}{2}I+i\,\, sin\frac{\theta}{2}\left[\begin{array}{cc}
1 & 0\\
0 & -1
\end{array}\right], \label{eqn2}
\end{equation}


\begin{equation}
U_{\theta} = cos\frac{\theta}{2}I+i sin\frac{\theta}{2}Z, \label{eqn3}
\end{equation} 
where I and Z are identity and Pauli operators, respectively.

\begin{equation}
|d\rangle = U_{\theta}|\psi\rangle = \alpha|0\rangle+\beta|1\rangle, \label{eqn4}
\end{equation}

where $|\psi\rangle$ is the quantum state sent by Alice to Bob, this $|\psi\rangle$ is converted  to a damaged quantum state $|d\rangle$, by applying $U_{\theta}$ transformation. Hence the damaged quantum state ($|d\rangle$) received by Bob is not same as input quantum state $|\psi\rangle$. This change occurs due to noisy quantum channel. At this stage it is necessary to apply an effective quantum error correction scheme which is known as pilot based quantum error correction scheme.
 
 \section{Quantum Error Correction Technique for Quantum-Based Satellite Communication} 
 Here we follow the redundancy-free error correction scheme \cite{bacsardi2009solutions, khalighi2014survey}. We represent the size of pilot and data qubits by $r$ and $n$, respectively, as shown in Fig. \ref{fig2}. The input pilot  and data qubits are $I_{p}$ and $I_{d}$, respectively.  Also, the output pilot and data qubits are  $O_{p}$ and $O_{d}$, respectively.


\begin{figure}[h]
\centering
\includegraphics[width=0.45\textwidth]{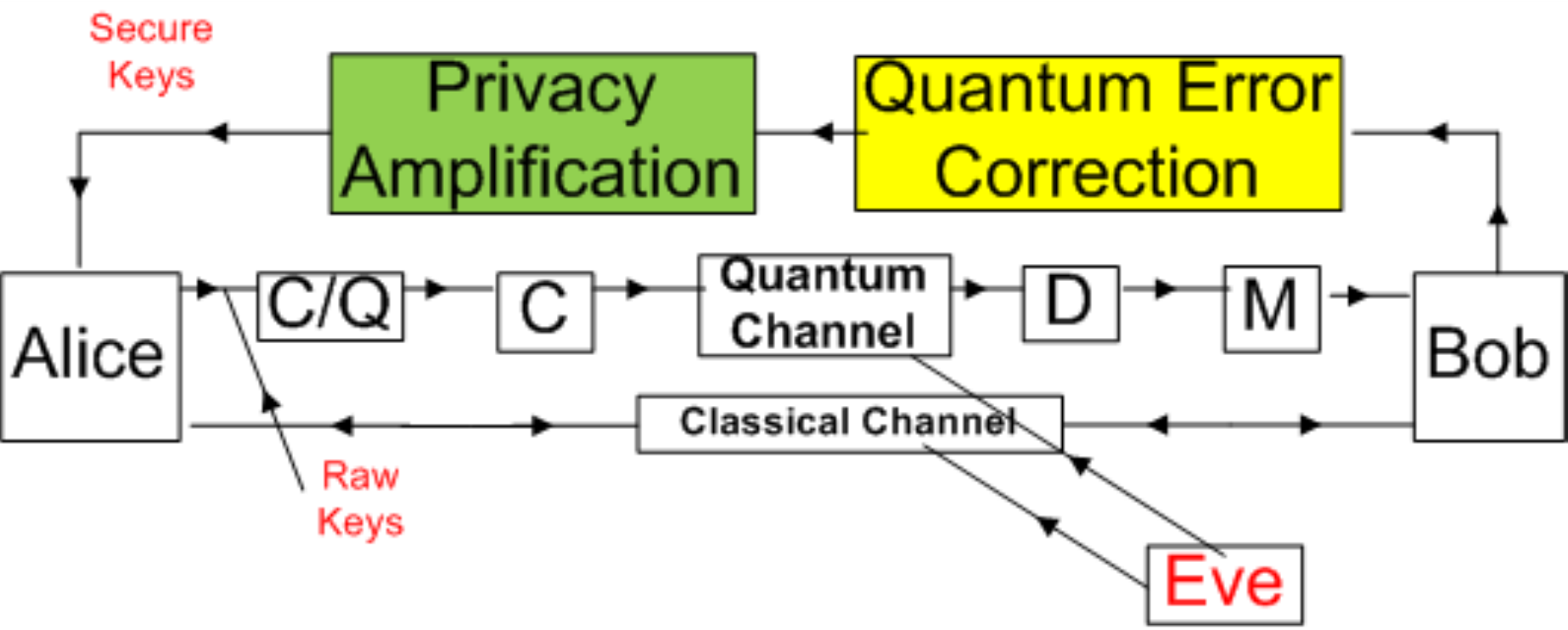}
\caption[1]{Quantum  communication procedure. Here $C$, $D$ and $M$ represents encoding, decoding and measurement operations, respectively. $C/Q$ represents classical to quantum conversion.} \label{fig1}
\end{figure}

In quantum-based satellite communication we assume that $U_{\theta}$  is constant for a time interval T. Hence, considering its effect, we receive the output pilot qubit state as 
\begin{equation}
O_{p} = U_{\theta}I_{p}U^{\dagger}_{\theta} = Tr|\theta\rangle\langle\theta|, \label{eqn5}
\end{equation}

\begin{figure}[h]
\centering
\includegraphics[width=0.50\textwidth]{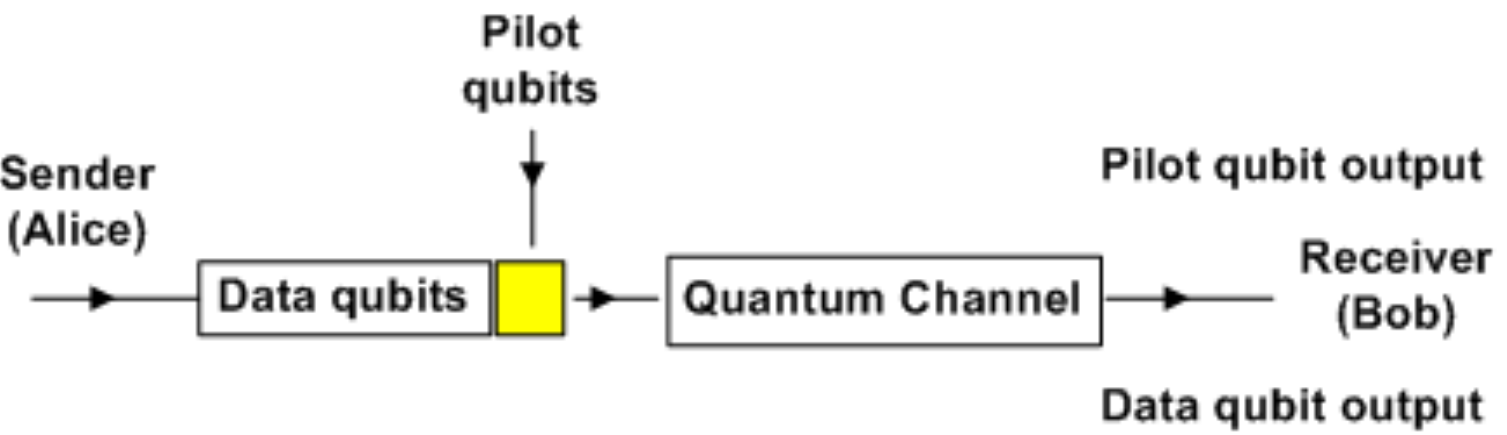}
\caption[1]{Quantum  error correction scheme.}, \label{fig2}
\end{figure}
\begin{figure}[h]
\centering
\includegraphics[width=0.48\textwidth]{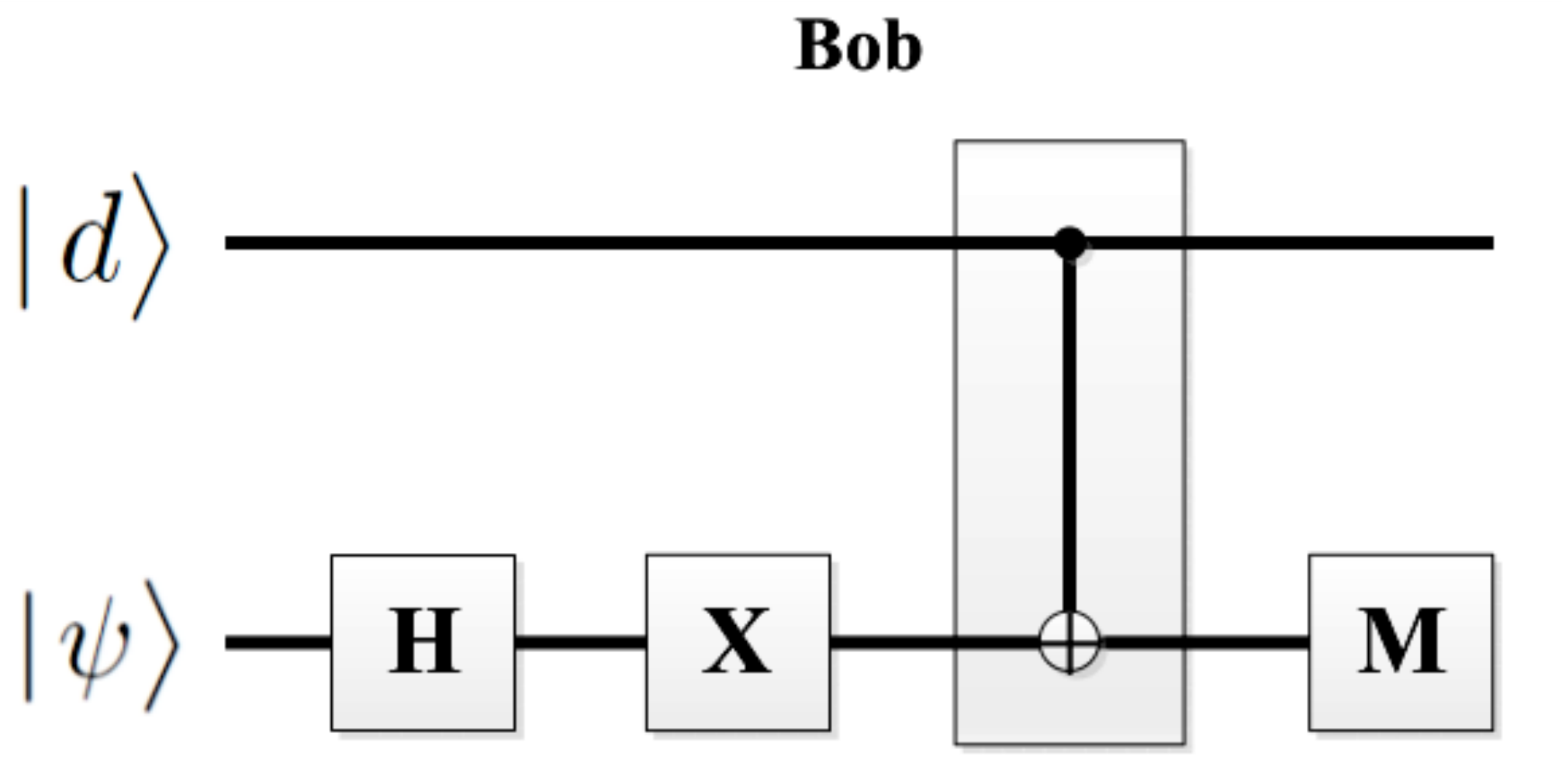}
\caption[1]{Quantum error correction circuit. Here H, X, M represent Hadamard gate, Pauli X gate, and measurement operator, respectively.} \label{fig4}
\end{figure}

\begin{equation}
|\theta\rangle\langle\theta|=(cos\frac{\theta}{2}|0\rangle+i sin\frac{\theta}{2}|1\rangle)(cos\frac{\theta}{2}\langle 0|+i sin\frac{\theta}{2}\langle 1|). \label{eqn6}
\end{equation}
We consider the redundancy-free model \cite{bacsardi2009solutions, khalighi2014survey}  in the quantum-based satellite communication to correct the transmitted qubits and obtain the improved keys, as shown in Fig. \ref{fig1}.
\begin{equation}
|d\rangle\otimes|\theta\rangle=\frac{1}{\sqrt2}[U^{\dagger}_{\theta}|d\rangle\otimes|0\rangle+ U_{\theta}|d\rangle\otimes|1\rangle]. \label{eqn7}
\end{equation}
 In the above equation, $|d\rangle$ is the damaged qubit (known as control qubit as shown in Fig. \ref{fig4}), and $|\theta\rangle$ is the error-correction rotation state (known as a target qubit as shown in Fig. \ref{fig4}). In addition to this, $\otimes$ is known as Kronecker product or tensor product.  From Eqs. \ref{eqn4} and \ref{eqn7}, Bob corrects the error
 
\begin{equation}
U^{\dagger}_{\theta}|d\rangle = U^{\dagger}U_{\theta}|\psi\rangle = |\psi\rangle, \label{eqn8}
\end{equation}
\begin{equation}
U_{\theta}|d\rangle= U_{\theta}U_{\theta}|\psi\rangle \neq |\psi\rangle. \label{eqn9}
\end{equation}


As shown in Fig. \ref{fig4}, Bob applies his measurement strategies on lower wire to get the qubit state $|0\rangle$ or  $|1\rangle$. From Eq. \ref{eqn7}, the correct key can be seen to be obtained with the probability half. This therefore implies that for the efficacy of the scheme multiple qubit should be used.

Now, assuming that Bob uses an $\xi$-length  qubit string for correcting  $|d\rangle$, 
\begin{equation}
|\theta\rangle_{m} = \otimes^{\xi}_{i=1}\,\ |2^{i-1}\theta\rangle, \label{eqn10}
\end{equation}
\begin{equation}
p_{e} = (1/2)^{\xi}. \label{eqn11}
\end{equation}
$p_{e}$ denotes the failure probability of Bob's error correction.
Bob considers  $|d\rangle$ as the control qubit, and   $\otimes^{\xi}_{i=1}\,\ |2^{i-1}\theta\rangle$ as the target qubit. Thus Bob obtains the updated keys as
\begin{eqnarray}
|d\rangle\otimes|\theta\rangle_{m}=|d\rangle\otimes^{\xi}_{i=1}|2^{i-1}\theta\rangle \\ \nonumber
=\frac{1}{\sqrt{2^\xi}}(\sqrt{2^{\xi}-1}U^{(2^{\xi}-1)\dagger}_{\theta}|d\rangle\otimes|correct\rangle\\ \nonumber
+U_{\theta}|d\rangle\otimes|error\rangle). \label{eqn12}
\end{eqnarray}
 For achieving an improved quantum security and efficiency,  we can use multiple data and multiple pilot qubit strings. Let $n$ and $\xi$ represent size of the data and pilot qubits, respectively. Let the received damaged multiple qubits by Bob are $|d_{1}\rangle \otimes |d_{2}\rangle \otimes |d_{3}\rangle \otimes  |d_{4}\rangle ...\otimes |d_{n}\rangle$. Hence, Bob takes multiple pilot qubits $|\theta\rangle \otimes |2 \theta\rangle \otimes |3 \theta\rangle \otimes |4 \theta\rangle ....\otimes |2^{\xi - 1}  \theta\rangle$ to correct the multiple  information carriers being communicated between earth-space stations.
\begin{eqnarray}
|d\rangle_{n}\otimes|\theta\rangle_{m} = \frac{1}{\sqrt{2^\xi}}(\sqrt{2^{\xi}-1}U^{(2^{\xi}-1)\dagger}_{\theta}|d\rangle_{n}\otimes|correct\rangle\\ \nonumber
+U_{\theta}|d\rangle_{n}\otimes|error\rangle), \label{eqn13}
\end{eqnarray}
\begin{equation}
|d\rangle_{n}= \sum^{2^{n}-1}_{k=0} D_{k}|k\rangle, \label{eqn14}
\end{equation}
where $D_{k}$ is the independent complex coefficient which satisfies $|D_{0}|^{2}+|D_{1}|^{2}+ --- + |D_{2^{n}-1}|^{2}=1$.\newline
 \newline
 In  detail,
\begin{eqnarray}
|d\rangle_{n}\otimes|\theta\rangle_{m}
=\frac{1}{\sqrt{2^\xi}}(\sqrt{2^{\xi}-1}U^{(2^{\xi}-1)\dagger}_{\theta}\sum^{2^{n}-1}_{k=0} D_{k}|k\rangle\otimes|correct\rangle\\ \nonumber
+U_{\theta}\sum^{2^{n}-1}_{k=0} D_{k}|k\rangle\otimes|error\rangle)\\ \nonumber
=\sum^{2^{n}-1}_{k=0}D_{k}\frac{1}{\sqrt{2^\xi}}(\sqrt{2^{\xi}-1}U^{(2^{\xi}-1)\dagger}_{\theta}|k\rangle\otimes|correct\rangle\\ 
+U_{\theta}|k\rangle\otimes|error\rangle). \nonumber \label{eqn15}
\end{eqnarray}
$p_{s} = 1 - (\frac{1}{2})^{\xi}$ represents success error correction probability which does not depend on  the length of the data qubit string. From Eq. \ref{eqn11}, we can say that in both the cases (single and multiple qubit data and pilot based schemes), the error correction probability is the same, as shown in Eq. \ref{eqn11}.\newline

From Fig. \ref{error correction prob}, it is clear that the success probability of the pilot error-correction increases with the increased number of pilot states. Following Fig. \ref{fig4} and Eq. \ref{eqn10}, the $\xi$ length pilot qubit strings can be generated from r pilot states $|\theta\rangle$:

\[\otimes_{i=1} ^{\xi} |2^{i-1}\theta\rangle = |\theta\rangle\otimes......\otimes|2^{\xi-1} \theta\rangle\]
\[r = 2^{\xi-2}(\xi-1)+2^{\xi-3}(\xi-2)+......+2^{2}.3+2^{1}.2+2^{0}.1+1+\xi-1 \]. \newline
The required number of pilot states and their success probability is shown in Table \ref{table:nonlinnn}.

 \begingroup
	    \setlength{\tabcolsep}{4pt} 
	    \renewcommand{\arraystretch}{3} 
	    \begin{table}[]
	    \caption{Required number of pilot qubits and \newline success probability for quantum error correction}
        \begin{tabular}{| c | c | c |}
	     \hline 
	     Length of the            &       Number of             &  Success probability       \\
	     pilot qubit ($\xi$)      &       pilot qubits (r)      &  ($p_{s}=1-(\frac{1}{2})^{\xi}$)\\     
	     \hline  \hline
	     $\xi$ =2 & r =3 & p = 0.75 \\
	    \hline
	    $\xi$ =3 & r =8 & p = 0.875 \\  
	    \hline 
	    $\xi$ =4 & r =21 & p = 0.9375 \\ 
	    \hline
	    $\xi$ =5 & r =54 & p = 0.96875 \\ 
	    \hline
	    $\xi$ =6 & r =135 & p = 0.984375 \\
	      \hline
	    \end{tabular}
	    \label{table:nonlinnn}
	    \end{table}
	     \endgroup
    
 	  	  \begingroup
 	  	 	  \setlength{\tabcolsep}{2.5pt} 
 	  	 	  \renewcommand{\arraystretch}{1.0} 
 	  	 	   \begin{table}[]
 	  	 	  	    \caption{ Calculated error correction data \newline for MEO (Medium-Earth Orbit satellite)}
 	  	 	  	      \centering
 	  	 	  	          \begin{tabular}{|c| c| c| c| c|}
 	  	 	  	       \hline
 	  	 	 Laser &   Raw qubits  & Maximal                   & Corrected qubits & D\\
 	  	 	 repetition rate &            & Transmittable qubits  & with $p_{s}$ = 0.96875 \\

 	  	 	  	    \hline
 	  	 	  	      10 GHZ & $5.10^{9}$ & $25.10^{4}$ & N = 249946 & 0.021 $\%$ \\
 	  	 	  	      \hline
 	  	 	  	       5 GHZ & $2.5.10^{9}$ & $12.5.10^{4}$ & N = 124946 & 0.04 $\%$ \\
 	  	 	  	       \hline
 	  	 	  	  1 GHZ & $5.10^{8}$ & $25.10^{3}$ & N = 29946 & 0.21 $\%$ \\
 	  	 	  	  \hline
 	  	 	  	    500 MHZ & $2.5.10^{8}$ & $12.5.10^{3}$ & N = 12446 & 0.43 $\%$ \\
 	  	 	  	    \hline
 	  	 	  	    100 MHZ & $5.10^{7}$ & $25.10^{2}$ & N = 2446 & 2.16 $\%$ \\
 	  	 	  	      \hline
 	  	 	  	       \end{tabular}
 	  	 	  	     \label{table:nonlin}
 	  	 	  	     \end{table}
 	  	 	  	    \endgroup    
 	  	     
\section{Quantum-based satellite communication:  performance analysis}
 Here we obtain secret key transmission,  assuming a Low Earth Orbit (LEO) satellite-ground quantum communication. Assuming the free space channel is constant during  time interval $T$ = 0.5s, we use a laser  with a repetition rate $f = 100MHz$ \cite{villoresi2008experimental, hughes2002practical, gyongyosi2011pilot, bradler2010trade, brandao2012does, gyongyosi2012quantum, gyongyosi2012classical}. The raw key qubits produced are given as
\begin{figure}[h]
\centering
\includegraphics[width=0.50\textwidth]{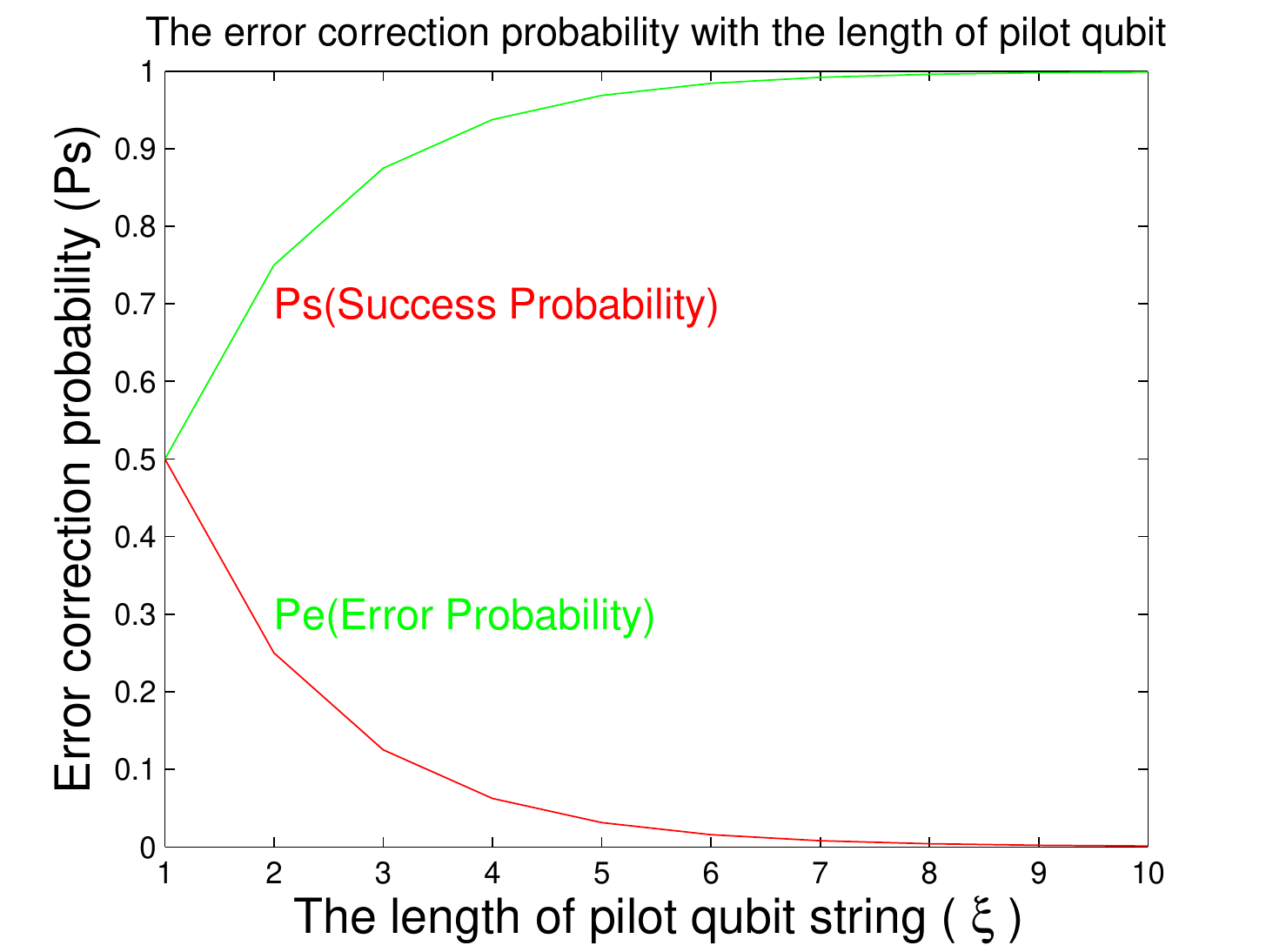}
\caption[1]{Performance comparison between error correction probability and length of the pilot qubit.} \label{error correction prob}
\end{figure}

\begin{figure}[h]
\centering
\includegraphics[ width= 0.53\textwidth]{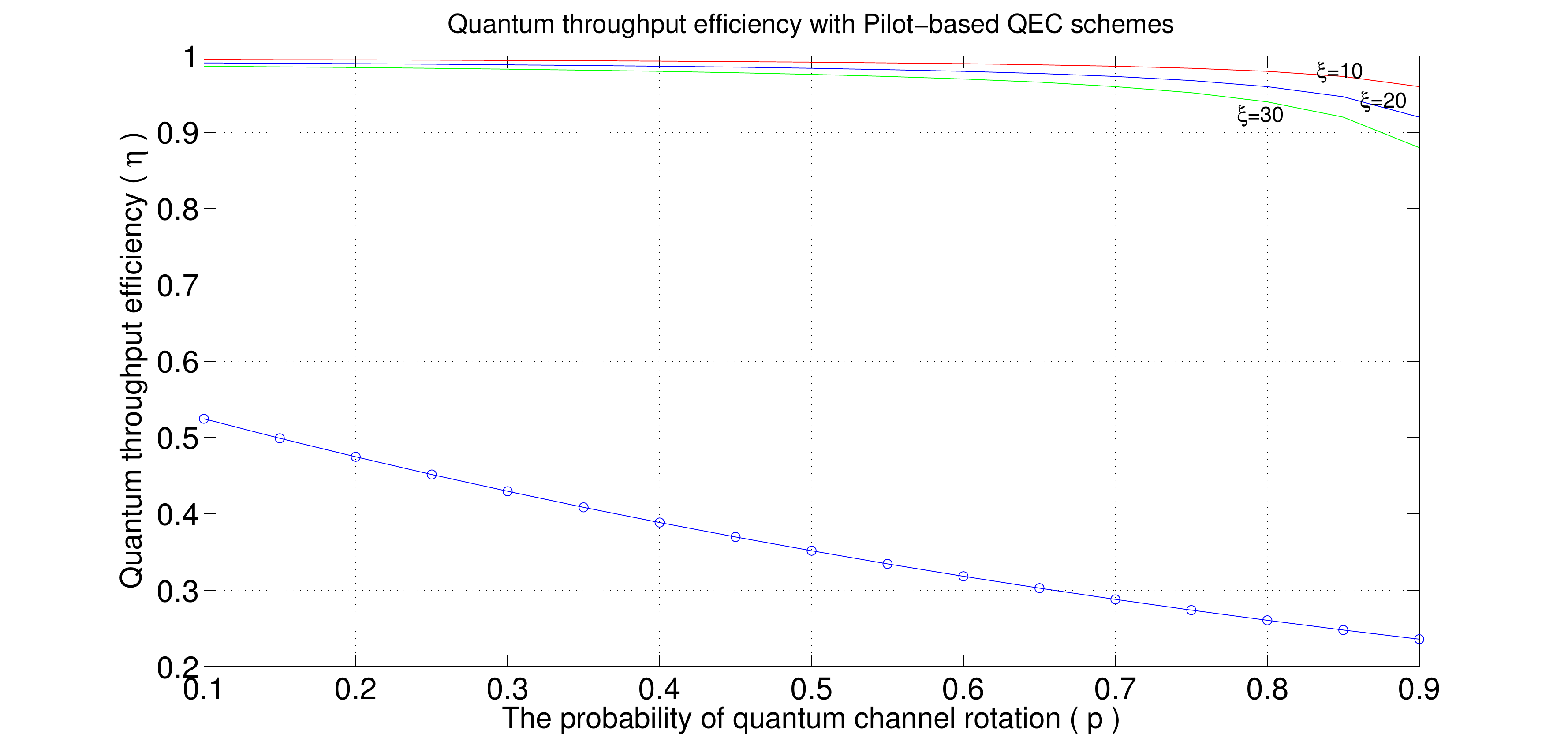}
\caption[1]{Quantum throughput efficiency comparison between classical cascade QEC (lower blue line) and pilot-based QEC (upper three lines)  for N = 2500, where $\xi$ = 10, 20 and 30 are the length of the pilot qubits.} \label{quantum through2500}
\end{figure}

\begin{equation}
R_{r}=f\times T=0.5\times10^{8}=5\times10^{7} qubits/s. \label{eqn16}
\end{equation}
We use an attenuation $\delta = 5 \times 10^{-5}$,  produced by telescope size and detector dark count rate. The size of the data and the pilot qubits are (n + $\xi$). Secure keys transmitted during T are 
\begin{equation}
R_{t}=R_{r}\times\delta=2500\,\ qubits/s. \label{eqn17}
\end{equation}
Efficiency of the quantum error correction scheme depends on the time T, for which the quantum channel is considered in a stationary state. The satellite or orbit speed decides the changes in angle $\theta$. Variation in time T depends on satellite orbit system (varies with low and high orbit systems) \cite{bonato2009feasibility}. In case of Geostationary Earth Orbit (GEO),  slow variation in rotation angle takes place, hence the quantum channel  sustains the stationary state for longer duration T, as compared to other low orbit satellite systems such as LEO (Low Earth Orbit) and MEO (Medium Earth Orbit) \cite{kurtsiefer2002quantum}. In addition to this,  performance varies with low to high orbit systems, variations in rotation angle and time parameter T.\newline

Quantum error correction data calculated for MEO satellite is shown in Table \ref{table:nonlin}. The calculation parameters for MEO satellite are as follows T = 0.5 sec.,  attenuation ($\delta) = 5 \times 10^{-5}$. Raw or generated qubits and total transmitted qubits ($N = n+r$) during time interval $T$ are calculated as in the case of LEO satellite. One more performance parameter is redundancy ($D$) which is calculated as \[D = \frac{r}{r+n} = \frac{r}{N}, \]  where value of  $r$ can be selected as per decoding success probability. The values shown in Table \ref{table:nonlin} can be used for real field applications such as quantum-based MEO satellite communication. \newline
\newline
Quantum throughput efficiency $\eta$ is
\begin{equation}
\eta = 1-\left[\frac{\frac{1}{1-p}\times\xi}{N}\right] \label{eqn18},
\end{equation}
where the probability corresponding to quantum channel rotation is $p$, length of the pilot-qubit in quantum error correction is $\xi$ and length of the total transmitted qubits  using quantum channel is $N$.
 Fig. \ref{quantum through2500} shows that the pilot-based quantum error correction in quantum-based satellite communications outperforms the Cascade quantum error correction scheme (see lower blue line) in terms of quantum throughput efficiency, because the pilot-based quantum error correction can obtain a high error correction probability (Fig. \ref{error correction prob}) with a small fraction of pilot qubit in total transmitted qubits. 
\section{CONCLUSIONS}
We use an effective pilot-based quantum error correction to remove the effects produced by the noisy quantum channel in quantum-based  satellite networks. The simulation results show that the pilot-based quantum error correction scheme in quantum key distribution satellite networks outperforms the classical Cascade quantum error correction scheme in terms of quantum throughput efficiency, which makes  quantum-based satellite communication feasible.

\end{document}